\documentclass[a4paper,twoside,twocolumn,english,amssymb,aps,prb,reprint,floats,showkeys,showpacs]{revtex4-1}
\usepackage{graphicx,amsmath}
\usepackage{epstopdf}
\usepackage{color}

\usepackage{tikz}
\newcommand{\dittotikz}{%
    \tikz{
        \draw [line width=0.12ex] (-0.2ex,0) -- +(0,0.8ex)
            (0.2ex,0) -- +(0,0.8ex);
        \draw [line width=0.08ex] (-0.6ex,0.4ex) -- +(-16.5em,0)
            (0.6ex,0.4ex) -- +(16.5em,0);
    }%
}

\begin{document}
\title{Linear scaling approach for atomistic calculation of excitonic
properties of 10-million-atom nanostructures}
\author{Piotr T. R\'o\.za\'nski}
\altaffiliation{present address: College of Inter-Faculty Individual Studies
in Mathematics and Natural Sciences (MISMaP), University of Warsaw,
ul. Banacha 2c, 02-097 Warszawa}
\author{Micha\l{} Zieli\'nski}
\email{mzielin@fizyka.umk.pl}
\affiliation{Institute of Physics, Faculty of Physics, Astronomy and
Informatics, Nicolaus Copernicus University, Grudziadzka 5, 87-100 Torun, Poland}

\begin{abstract}
Numerical calculations of excitonic properties of novel nanostructures, such
as nanowire and crystal phase quantum dots, must combine atomistic accuracy
with an approachable computational complexity.
The key difficulty comes from the fact that excitonic spectra details arise from
atomic-scale contributions that must be integrated over a large spatial
domain containing a million and more of atoms.
In this work we present a step-by-step solution to this problem: combined
empirical tight-binding and configuration interaction scheme that unites
linearly scaling computational time with the essentials of the atomistic
modeling.  We benchmark our method on the example of well-studied
self-assembled InAs/GaAs quantum dot.  Next, we apply our atomistic approach
to crystal phase quantum dots containing more than 10 million atoms.
\end{abstract}
\maketitle

Accurate calculations of excitonic properties of semiconductor quantum
dots\cite{michler} must fulfill two apparently contradictory conditions.
The first constraint is the necessity of handling computational domains
containing millions of atoms.\cite{bester-zunger-nair,alloyed-qd}
The second requirement is the atomistic accuracy necessary for the accurate
description of the excitonic
spectra.\cite{jaskolski-zielinski-prb06,bryant-prl,singh-bester-eh,ediger-zunger-aufbau}
First principles modeling such as the density functional theory\cite{dftadv}
or GW-Bethe-Salpeter-equation approach\cite{onida}
for self-assembled\cite{arek-book} or nanowire\cite{bjork} quantum dots is
still beyond the reach of modern computers.
Continuous matter approaches like the effective mass approximation or
$k\cdot p$ method\cite{stier-grundmann-bimberg,kp-schliwa}
have demonstrated the capability of describing main features in QDs
spectra;\cite{arek-book,bimberg-book} however, these methods are restricted
by the resolution on the scale of a unit cell.\cite{delerue-book}
One of the key examples is the bright exciton
splitting,\cite{bayer-eh,takagahara,kadantsev-eh} where sophisticated
multi-band approaches are needed to accommodate for the correct symmetry of
the underlying crystal
lattice.\cite{kp14,kp14more,karlsson,zielinski-tallqd}
Apart from symmetry issues, even the most elaborate continuous matter
tactics cannot fully account for effects of alloying,\cite{alloyed-qd}
lattice randomness,\cite{luo-zunger}, interface
effects,\cite{niquet-band-offset,zwan} or crystal phase
symmetry\cite{nika-cpqd,our-cpqd} for which a truly atomistic approach is
needed.

Practical methods, capturing atomistic structure of quantum dots and their
surrounding matrix, include two semi-empirical approaches: the empirical
tight-binding\cite{leung-whaley-prb1997,lee-johnsson-prb2001,santoprete,usman1,schulz-schum-czycholl,zielinski-prb09,zielinski-including,zielinski-vbo}
and the empirical pseudopotential
method.\cite{Zunger-EPM2,bester-zunger-p-shell,gong-prb77}
Typical calculations using these approximate approaches involve three
subsequent
stages:\cite{Zunger-EPM1,sheng-cheng-prb2005,korkusinski-zielinski-hawrylak-jap09,bryant-prb}
a) calculation of equilibrium position of constituent atoms, b) calculation
of quasielectron and quasihole states, and c)
inclusion of excited quasiparticles interactions via screened Coulomb
potential, solved by some sort of configuration interaction (exact
diagonalization) method.

Although the stages of finding positions (strains) and quasi-particle
calculations are often far from trivial, the key computational issue for
approaches aiming to quantitatively describe nanosystem spectra is the
ultimate step: the many-body calculation.  This part of calculation is of
particular importance as in a typical quantum dot photoluminescence
experiment one does not observe the single-particle spectra,
but rather many-body spectra of charged and neutral (multi-) excitonic
complexes.\cite{michler}
Therefore in a realistic calculation the single particle part of the
computation must be followed by a many-body calculation.
Similarly to post-Hartree-Fock methods of quantum
chemistry,\cite{szabo-ostlund} the many-body calculation following the
single-particle part is often the most time demanding stage of the entire
empirical computations sequence.
In particular, as a necessary prerequisite for the many-body calculation
typically numerous two-particle integrals must be first calculated using
quasiparticle functions.
In the tight-binding (TB) approach, for $N$ atoms, $A$ basis functions per atom,
and $M$ quasiparticle states there are $O(N^4A^4M^4)$ two-particle
integrals, and the computational effort will thus formally scale as the
fourth power in number of atoms, what leads to a practically untractable
problem for typical quantum dots with numbers of atoms exceeding a million.

In this paper we present an approximated method for efficient order-$N$
calculation of screened Coulomb and exchange integrals within the
tight-binding framework.
Apart from linear scaling our method accounts for terms that are customarily
neglected in a typical tight-binding calculation of Coulomb matrix elements.
We test our approach on the example of million-atom InAs/GaAs lens-shape
self-assembled quantum dots and finally apply our approach to
multi-million-atom crystal phase quantum dots.

\section{Coulomb matrix elements}
At the moment one of the most successful empirical computational procedures
is the empirical pseudopotential method.\cite{Zunger-EPM2,Zunger-EPM1}
This approach uses ``realistic'' empirical pseudopotentials that reproduce
correct band gaps and effective masses.
Those potentials can be accommodated with a relatively small plane wave
basis set
and are used to obtain the single-particle eigenvectors as solution of the
Schr\"{o}dinger equation with an auxiliary basis set of strained Bloch
function of the underlying bulk.\cite{zunger}
Whereas conceptually straightforward, the practical implementation of the
empirical pseudopotential method for quantum dots has been so far limited
only to A.~Zunger and co-workers.

On the contrary, numerous theoretical groups utilize different flavors of
the empirical tight-binding method.
The latter approach origins from the Slater-Koster scheme of orthogonal
tight-binding,\cite{slater-koster} whereas Hamiltonian matrix elements are
given in terms of several empirical constants\cite{harrison-big-book}
determined to reproduce bulk properties such as effective masses, bulk
deformation potentials and gaps at different points of the Brillouin
zone.\cite{jancu}
Fitting process is usually far from trivial\cite{boykin-fit}; however, for a
wide family of materials, there are reliable and accurate tight-binding
parameter sets available in the
literature.\cite{jancu,boykin-fit,vogl,boykin-diagonal-shift,Sapra,Viswanatha}

The tight-binding method applied to a typical nanowire or self-assembled
quantum dot problem produces a Hamiltonian matrices of dimensions extending
$10^6$;
however, due to the nearest-neighbors approximation the Hamiltonian matrix
is sparse.
Then, for selected single particle states (close to energy band gap) the
eigenproblem can be solved efficiently using e.g.  Krylov iterative methods,
such as Lanczos\cite{lanczos} or Arnoldi\cite{arnoldi} algorithm, therefore
resulting in a linear scaling of computational time with respect to the
number of atoms in a computational domain.
Like many other numerical approaches the empirical tight-binding benefits
considerably from the parallelization that further reduces the overall
computation time.\cite{alloyed-qd}

The tight-binding or linear combination of atomic orbitals (LCAO) wave
function is given as:
\begin{align}
\psi(\vec{r})=\sum_{\vec{R}}^N\sum_{\alpha}^Ab_{\vec{R}\alpha}\left|\vec{R}\alpha\right>=\sum_{\vec{R}}^N\sum_{\alpha}^Ab_{\vec{R}\alpha}\phi_{\vec{R}\alpha}(\vec{r}-\vec{R})
\label{TB}
\end{align}
where the summation goes over all ($N$) atoms $\vec{R}$ and all ($A$) over
atomic orbitals $\alpha$ centered on a given atom, whereas
$b_{\vec{R}\alpha}$ are basis expansion coefficients.
Assuming a statically screened\cite{onida} Coulomb interaction, the Coulomb
matrix elements $V_{ijkl}$ are given by:~\cite{zielinski-prb09}
\begin{align}
V_{ijkl}=\int \int \psi_i^*(\vec{r_1}) \psi_j^*(\vec{r_2})
\frac{e^2}{\epsilon(\vec{r_1},\vec{r_2})\left|\vec{r_1}-\vec{r_2}\right|}
\psi_k(\vec{r_2})\psi_l(\vec{r_1})
\label{coulomb-general}
\end{align}
where $\epsilon(\vec{r_1},\vec{r_2})$ is the position-dependent dielectric
function and $\psi$'s are single-particle, electron or hole, wave functions.
By substituting single-particle wave functions in form of Eq.~\ref{TB} into
Eq.~\ref{coulomb-general} one obtains:
\begin{multline}
V_{ijkl}=
\sum_{\vec{R_1}\alpha_1}
\sum_{\vec{R_2}\alpha_2}
\sum_{\vec{R_3}\alpha_3}
\sum_{\vec{R_4}\alpha_4}
b_{\vec{R_1}\alpha_1}^{i*}
b_{\vec{R_2}\alpha_2}^{j*}
b_{\vec{R_3}\alpha_3}^{l}
b_{\vec{R_4}\alpha_4}^{k} \times \\
\omega(\vec{R_1}\alpha_1,\vec{R_2}\alpha_2,\vec{R_3}\alpha_3,\vec{R_4}\alpha_4),
\label{CME}
\end{multline}
where four-fold summation goes over all atomic positions and orbitals and
{\footnotesize
\begin{align}
&\omega(\vec{R_1}\alpha_1,\vec{R_2}\alpha_2,\vec{R_3}\alpha_3,\vec{R_4}\alpha_4)\equiv\\
\nonumber
&\int \int \phi_{\vec{R_1}\alpha_1} \left(\vec{r_1}\right)
\phi_{\vec{R_2}\alpha_2} \left(\vec{r_2}\right)
\frac{e^2}{\epsilon\left(\vec{r_1},\vec{r_2}\right)\left|\vec{r_1}-\vec{r_2}\right|}
\phi_{\vec{R_3}\alpha_3} \left(\vec{r_2}\right) \phi_{\vec{R_4}\alpha_4}
\left(\vec{r_1}\right)
\end{align}
}
is an integral calculated in a basis of tight-binding (``atomic'') orbitals.

If treated directly, this procedure would result in $O(N^4A^4)$ terms
(atomic integrals) constituting one quasi-particle Coulomb matrix element,
where $N$ is the number of atoms in the domain (typically $N\sim10^6$ for
self-assembled and nanowire quantum dots) and $A$ is number of
(spin-)orbitals associated with each of atoms (e.g.  $20$ for the
$sp^3d^5s^\star$ TB model\cite{zielinski-including}).
To further complicate matters, practical calculation demands computation not
of one, but numerous electron-electron, hole-hole, and electron-hole Coulomb
matrix elements calculated using Eq.~\ref{CME}.
For example, in a typical quantum dot calculation involving $12$ electron
states and $12$ hole states (including spin), the total number of these
quasi-particle Coulomb matrix elements reaches $10^5$, whereas
in certain situations\cite{zielinski-tallqd} number of quasi-particle states
($M$) used for the calculation must be further extended.
Either way, the overall computation time scales as $O(N^4A^4M^4)$.

Typically, this formidable problem is resolved by utilizing a series of
approximations,\cite{schulz-schum-czycholl,zielinski-prb09} including
neglect of three- and four-center integrals
$\omega$.\cite{lee-johnsson-prb2001}
Further approximations involve multi-pole expansion of single integral
$\omega$ and retaining monopole-monopole contributions
only.\cite{schulz-schum-czycholl,zielinski-prb09}

Finally one gets an approximate form of Coulomb matrix
elements:\cite{zielinski-prb09}
\begin{align}
\resizebox{.9\hsize}{!}{$
V_{ijkl}=
\sum_{\vec{R_1}}\sum_{\vec{R_2}\neq\vec{R_1}}
\left[\sum_{\alpha_1}
b_{\vec{R_1}\alpha_1}^{i*}b_{\vec{R_1}\alpha_1}^{l}\right]
\left[\sum_{\alpha_2}
b_{\vec{R_2}\alpha_2}^{j*}b_{\vec{R_2}\alpha_2}^{k}\right]
\frac{e^2}{\epsilon\left|\vec{R_1}-\vec{R_2}\right|}
+\nonumber $} \\
\resizebox{.9\hsize}{!}{$
\sum_{\vec{R_1}}
\sum_{\alpha_1\alpha_2\alpha_3\alpha_4}
b_{\vec{R_1}\alpha_1}^{i*}b_{\vec{R_1}\alpha_2}^{j*}
b_{\vec{R_1}\alpha_3}^{k}b_{\vec{R_1}\alpha_4}^{l}
\omega(\vec{R_1}\alpha_1,\vec{R_1}\alpha_2,\vec{R_1}\alpha_3,\vec{R_1}\alpha_4),
$}
\label{coulomb-special}
\end{align}
where the first term is the long-range, bulk-screened, contribution to the
two-center integral built from the monopole-monopole
interaction\cite{Franceschetti,Goupalov} of two charge densities localized
at different atomic sites.
The second term is the on-site unscreened part, calculated by direct
integration using atomic
orbitals.\cite{lee-johnsson-prb2001,leung-whaley-prb1997}
This approach is justified by the fact that the screening (Thomas-Fermi)
radius ($\approx$~2--4~\AA) is on the order of a bond length
\cite{lee-johnsson-prb2001,delerue-book} resulting in nearly bulk screening
of off-site (long-range) terms and limited screening of on-site
(short-range) terms contribution.
A potential problem arising here is the choice of atomic basis used for
calculation of $\omega$ on-site integrals.  The generally important matter
of basis dependence will be discussed later in more detail.

The second summation in Eq.~\ref{coulomb-special} over on-site terms can be
further simplified by neglecting exchange terms and multiple $\omega$'s
replaced by a single on-site contribution:\cite{schulz-schum-czycholl}
\begin{align}
V_{ijkl}=
\sum_{\vec{R_1}}\sum_{\vec{R_2}}
\left[\sum_{\alpha_1}
b_{\vec{R_1}\alpha_1}^{i*}b_{\vec{R_1}\alpha_1}^{l}\right]
\left[\sum_{\alpha_2}
b_{\vec{R_2}\alpha_2}^{j*}b_{\vec{R_2}\alpha_2}^{k}\right]
V_{\vec{R_1}\vec{R_2}}, \nonumber\\
V_{\vec{R_1}\vec{R_2}}=
\begin{cases}
    \frac{e^2}{\epsilon\left|\vec{R_1}-\vec{R_2}\right|},& \vec{R}\neq\vec{R'},\\
    U_{\vec{R}},              & \vec{R}=\vec{R'},
\end{cases}
\label{coulomb-special-simple}
\end{align}
where the on-site atomic contribution $U_{\vec{R}}$ can be calculated
(estimated) using different
approaches.\cite{schulz-schum-czycholl,delerue-book}
Whereas relatively uncomplicated, Eq.~\ref{coulomb-special} and
Eq.~\ref{coulomb-special-simple} give reasonable agreement with the
experiment and other computational
approaches.\cite{zielinski-including,zielinski-vbo,schulz-schum-czycholl}
The above procedures (Eq.~\ref{coulomb-special} and
Eq.~\ref{coulomb-special-simple}) give also far more approachable $O(N^2)$
scaling of a single Coulomb matrix element calculation rather than
impractical $O(N^4)$ of the straightforward method (Eq.~\ref{CME}).
Further reduction of the computational time (but not the scaling factor) can
be achieved by utilizing multi-scale\cite{zielinski-multiscale} approaches
(i.e.  using smaller domains at different stages of the computation).
While not affecting the scaling properties, the usage of numerical libraries
such as BLAS can further reduce the computation time by avoiding time
consuming recalculation\cite{sheng-cheng-prb2005} of intermediate terms.

Apart from the advantages, the above methods (Eq.~\ref{coulomb-special} and
Eq.~\ref{coulomb-special-simple}) reveal two apparent issues.  The first is
a non-linear $O(N^2)$ scaling of the computational time leading to exploding
computational complexity for domains containing millions of atoms.
The second problem is an possible inaccuracy related to the two-center and
the monopole-monopole approximations.

\section{Wave-function reconstruction}
In what follows we present a numerical method that effectively addresses
above issues.
Let us starts by reiterating that the LCAO representation of the
tight-binding wavefunction leads to a non-linear scaling of the two-particle
Coulomb matrix elements calculation with respect to the number of atoms in
the domain.
Therefore, after TB stage of calculation, we convert TB eigenstates from a
typical LCAO wave-functions to a real-space tight-binding wave-function
representations.
This stage is achieved by an introduction of a three dimensional, uniform
real-space grid with complex values.
At each point of the spatial grid we use Eq.~\ref{TB} (assuming particular
basis set) to calculate the wave-function values.
We repeat this procedure and process in this way each of several lowest
electron and hole functions obtained by the tight-binding procedure,
obtaining the real space grid representation for each of the considered
functions.
The resolution of the grid, common for all functions, is determined by
convergence studies and will be discussed later.
We denote this stage of calculation as the ``wave-function grid
reconstruction'' or simply ``wave-function reconstruction''.
The benefits of this transformation will become apparent soon.
Later we will discuss in detail the effect of the basis choice used in
Eq.~\ref{TB}.

The computational cost of the straightforward single wave-function
reconstruction using Eq.~\ref{TB} is proportional to number of atoms $N$ and
the number of grid points $P$ resulting in $O(NP)$ computation time for the
single quasi-particle function reconstruction.  For accurate computations
and high grid resolution, $P$ should clearly be proportional (even
exceeding) than $N$ leading to the unfavorable $O(N^2)$ scaling at this
stage of calculation.
However, let us notice that in an orthogonal tight-binding method
it is legitimate to assume that for any point in space $\vec{r}$ the value
of the tight-binding wave-function $\psi$ can be replaced by a contribution
from several atomic neighbors only lying within a certain ``cut-off'' radius
$R_{cut}$:
\begin{multline}
\psi\left(\vec{r}\right)=\sum_{\vec{R}}^N\sum_{\alpha}^Ab_{\vec{R}\alpha}\phi_{\vec{R}\alpha}(\vec{r}-\vec{R})
\\
\approx\sum_{\left|\vec{r}-\vec{R}\right|<R_\text{cut}}\sum_{\alpha}^Ab_{\vec{R}\alpha}\phi_{\vec{R}\alpha}(\vec{r}-\vec{R})
\label{tb-cut-off}
\end{multline}
More strictly, rather than due to orthogonality, this assumption originates
from atomic orbitals spatial locality, i.e.  relatively small spatial extent
of atomic orbitals when compared to dimensions of the entire system.
Similar assumptions are typically utilized in the context of other
linear-scaling approaches,\cite{linear-scaling} that rely heavily on the
use of strictly confined basis orbitals, i.e.  orbitals zeroing beyond a
certain radius.  One way of achieving this goal is the method of Sankey and
Niklewski\cite{sankey-niklewski} where the (pseudo-) atom is embedded within
a spherical box of finite radius.

Should $R_\text{cut}$ be much smaller then the system size then the number
of atoms ($N_\text{cut}$) within the cut-off radius: $N_\text{cut} \ll N$.
More importantly, for quantum dots embedded in bulk-like matrix,
$N_\text{cut}$ should be relatively constant and independent from $N$,
leading to $O(N)$ scaling of the tight-binding wave-function reconstruction
time.

A straightforward implementation of Eq.~\ref{tb-cut-off} would involve
iteration over grid points and therefore implicitly perform costly
calculations of distances between all grid points and all atoms.
Instead, in our implementation, we iterate over all atoms and update only
grid points within a neighborhood defined as a cube of edge length equal to
$2R_\text{cut}$.
This procedure can also be efficiently parallelized.

In case of semiconductor quantum dots above assumptions and exact value of
$R_\text{cut}$ will be verified later by numerical tests.
The size of the reconstruction grid should be equal to spatial system
dimensions increased in all directions by necessary margins equal to
$R_\text{cut}$.
We also note that we separately compute and store spin-up and spin-down
wave-function components at each grid point.
Once the reconstruction is finished, the wave-function is finally
renormalized on the grid.

With the TB wave-function given in real-space we can follow with the
calculation of Coulomb matrix elements using Eq.~\ref{coulomb-general}.
Straightforward substitution of real-space wave-functions into
Eq.~\ref{coulomb-general} and direct integration would however lead to
$O(P^2)\propto O(N^2)$ complexity.
Fortunately, this equation be conveniently evaluated in the reciprocal
space.
Eq.~\ref{coulomb-general} can be cast in a general form:\cite{oliveira}

\begin{equation}
V_{ijkl}=\int \int \rho_{il}(\vec{r}) \, G(\vec{r}-\vec{r'}) \, \rho_{jk}(\vec{r'}) \, dV dV',
\end{equation}

where $\rho_{il}=\psi^{i*}\psi^{l}$, $\rho_{jk}=\psi^{j*}\psi^{k}$ and
$G(\vec{r}-\vec{r'})$ is the screened Coulomb interaction.
Then, potential $V_{jk}$ calculated from quasi-density $\rho_{jk}$ is given
as:
\begin{equation}
V_{jk}(\vec{r})=\int G(\vec{r}-\vec{r'}) \, \rho_{jk}(\vec{r'}) \, dV'
\label{vjk}
\end{equation}
For a discrete case, assuming regular grid with the grid step $h$, we
obtain:
\begin{equation}
V_{jk}[\vec{r}] = \sum_{\vec{r'}} G[\vec{r}-\vec{r'}] \, \rho_{jk}[\vec{r'}]
\, h^3 = (G\ast\rho_{jk})[\vec{r}] \, h^3
\label{eq-Vjk}
\end{equation}
where $\vec{r}$ and $\vec{r'}$ are points on a discrete grid and
$(G\ast\rho_{jk})$ is a full convolution between quasidensity $\rho_{jk}$
and the screened Coulomb interaction $G(\vec{r}-\vec{r'})$, which is defined
on a three-dimensional domain twice as large as $\rho_{jk}$ (as it includes
both positive and negative shifts) in every dimension. Therefore, the
density grid must also be padded with zeros, resulting in $8P$ grid points
instead of $P$. Due to this padding, the full convolution
$(G\ast\rho_{jk})$ is equivalent to the circular convolution, which in turn
can be effectively computed using Fast Fourier Transform (FFT)
algorithm\cite{fft} in $O(P\log P)$ time.

Since we use FFT purely as a computational tool to calculate full (not
circular) convolution, we do not introduce any undesired periodicity to the
problem.  Details of this approach are given e.g.  in
Ref.~\cite{numerical-recipes}.  This technique is a standard in digital
signal processing for a discrete Fourier transform treatment of non-periodic
signals convoluted with a response function of a finite duration.
Otherwise, Eq.~\ref{eq-Vjk} would typically converge slowly with respect to
the grid (supercell) volume.  Different techniques including multi-pole
expansion\cite{makov-payne,oliveira,franceschetti-screening} or truncated
Coulomb interaction\cite{franceschetti-screening} would be typically
necessary to speed up this convergence.

The method we utilize seems not to be often used in the electronic structure
calculation, most likely due to the increased memory demand.  In our case
the size of the FFT domain is effectively doubled in all three directions.
As a practical benefit, we do not need to perform convergence tests with
respect to the domain size as Eq.~\ref{v_sum} gives exactly the same results
as the direct integration of Eq.~\ref{coulomb-general}.
In the latter part of the text we will additionally verify that statement by
performing numerical tests of the effective range of the Coulomb
interaction.
We also note that in other approaches the size of the supercell is often
much larger then the actual system size.\cite{franceschetti-screening}
Moreover, extending the FFT domain is not affecting the time needed for the
wave-function reconstruction, but only the (relatively short) time of the
FFT calculation.

$V_{ijkl}$ is finally given as a straightforward $O(P)\propto O(N)$
summation over all grid points:
\begin{equation}
V_{ijkl}=\sum_{\vec{r}} \rho_{il}[\vec{r}] \, V_{jk}[\vec{r}]h^3
\label{v_sum}
\end{equation}

For a given $jk$ pair of tight-binding (grid reconstructed) functions,
$V_{jk}$ and the FFT transform are calculated only once and are used to
calculate all resulting Coulomb matrix elements $V_{ijkl}$,
therefore avoiding costly recalculation of $V_{jk}$.

Our approach accounts for dielectric effects at different levels of
approximation.  For example, we can use any form of distance dependent
dielectric function, e.g.  a Thomas-Fermi model of Resta~\cite{resta}
customarily utilized in the empirical-pseudopotential method
calculations.\cite{franceschetti-screening} Apart from the
distance-dependence, we can account for the spatial dependence of the
dielectric medium.  Details of this latter approach will be discussed
elsewhere.  In the current paper $G$ is either taken as a Fourier transform
of Coulomb bulk-screened interaction or is given by a Thomas-Fermi model of
Resta.~\cite{resta}
Finally, we note that due to introduction of large spatial grid our method
demands substantially larger computer memory than the simple model
(Eq.~\ref{coulomb-special}).

\section{Lattice and the single particle spectra}
As discussed in the introduction, the calculation consists of several major
steps: first atomic positions are calculated.
For lattice mismatched system (such as InAs/GaAs quantum dots) to calculate
strain relaxed positions we use the atomistic valence force field (VFF)
approach of Keating.\cite{keating,martin}
This method is described in more detail in
Ref.\cite{pryor-zunger,saito-arakawa} and in our previous
papers.\cite{jaskolski-zielinski-prb06,zielinski-prb09,zielinski-including,zielinski-vbo}
We note here only that the VFF approach is a $O(N)$ method and can be
efficiently parallelized allowing for treatment of domains containing $10^8$
atoms.\cite{zielinski-vbo}
For InP crystal phase quantum dots, we neglect strain
effects.~\cite{nika-cpqd,our-cpqd}

Once the atomic positions are given, we use them to calculate single
particle energies with the empirical nearest-neighbor tight-binding model
that accounts for strain, spin-orbit interactions, crystal lattice symmetry,
and wurtzite crystal field splitting in case of zinc-blende/wurtzite mixed
crystal phase quantum dots.~\cite{zielinski-including,zielinski-vbo}

The single-particle tight-binding Hamiltonian for the system of $N$ atoms
and $m$ orbitals per atom can be written, in the language of the
second quantization, in the following form:
{\begin{multline}
  \hat{H}_{TB} =
  \sum_{i=1}^N \sum_{\alpha=1}^{m}
       E_{i\alpha}c_{i\alpha}^+c_{i\alpha}
+  \sum_{i=1}^N \sum_{\alpha=1,\beta=1}^{m}
       \lambda_{i\alpha,\beta}c_{i\alpha}^+c_{i\beta}\\
+   \sum_{i=1}^N \sum_{j=1}^{4} \sum_{\alpha,\beta=1}^{m}
       t_{i\alpha,j\beta}c_{i\alpha}^+c_{j\beta}
\end{multline}}
where $c_{i\alpha}^+$ ($c_{i\alpha}$) is the creation (annihilation)
operator of a carrier on the orbital $\alpha$ localized on the site
$i$, $E_{i\alpha}$ is the corresponding on-site (diagonal) energy, and
$t_{i\alpha,j\beta}$ describes the hopping (off-site, off-diagonal)
of the particle between the orbitals on (four) nearest neighboring sites.
Coupling to further neighbors is neglected, whereas
$\lambda_{i\alpha,\beta}$ (on-site, off-diagonal) accounts for the
spin-orbit interaction following the description given by
Chadi.\cite{chadi-so-in-tb}

For InAs/GaAs system we use tight-binding parameters set from
Ref.\cite{jancu} in $sp^3d^5s^\star$ parametrization.
More details of the $sp^3d^5s^\star$ tight-binding calculation were
discussed thoroughly in our earlier
papers.\cite{zielinski-including,zielinski-vbo}

\begin{table*}[]
\centering
\resizebox{0.9\textwidth}{!}{\begin{minipage}{\textwidth}
\caption{Tight-binding $sp^3s^\star$ parameters for InP zinc-blende and wurtzite phases. Right-hand ($11$) parameters are identical for both phases. }
\label{tb-params}
\begin{tabular}{|c|r|r|r|r|r|r|c|c|c|c|c|c|c|c|c|c|c|}
\hline
\multicolumn{1}{|l|}{} & \multicolumn{1}{c|}{$E_{sa}$} & \multicolumn{1}{c|}{$E_{pa}$} & \multicolumn{1}{c|}{$E_{p_za}$} & \multicolumn{1}{c|}{$E_{sc}$} & \multicolumn{1}{c|}{$E_{pc}$} & \multicolumn{1}{c|}{$E_{p_zc}$} & $E_{s^\star a}$             & $E_{s^\star c}$             & $V_{ss}$                     & $V_{xx}$                    & $V_{xy}$                    & $V_{sa,pc}$                 & $V_{sc,pa}$                 & $V_{s^\star a,pc}$               & $V_{s^\star c,pa}$               & $\lambda_a$               & $\lambda_c$               \\ \hline
ZB                     & -8.5274                       & 0.7677                        & 0.7677                          & -1.4826                       & 3.9407                        & 3.9407                          & \multicolumn{1}{r|}{8.2635} & \multicolumn{1}{r|}{7.0665} & \multicolumn{1}{r|}{-5.3614} & \multicolumn{1}{r|}{1.8801} & \multicolumn{1}{r|}{4.2324} & \multicolumn{1}{r|}{2.2265} & \multicolumn{1}{r|}{5.5825} & \multicolumn{1}{r|}{3.4623} & \multicolumn{1}{r|}{4.4814} & \multicolumn{1}{r|}{0.067} & \multicolumn{1}{r|}{0.392} \\ \hline
WZ                     & -8.4634                       & 0.8323                        & 0.8063                          & -1.4186                       & 4.0053                        & 3.9793                          & \multicolumn{11}{c|}{\dittotikz}                                                                                                                                                                                                                                                                                                       \\ \hline
\end{tabular}
\end{minipage}}
\end{table*}

For InP crystal phase quantum dots we have used Vogl et al.\cite{vogl}
$sp^3s^\star$ tight-binding parameters augmented to account for the
spin-orbit splitting (126 meV~\cite{chadi-so-in-tb}).
We have additionally modified these parameters to account for the increased
wurtzite's band gap ($1.474$~eV~\cite{de-pryor}), the valence band offset
($64.6$~meV~\cite{de-pryor}) between the wurtzite and zinc blende segments,
and the wurtzite crystal field splitting ($26$~meV~\cite{zhang-cpqd}).  The
tight-binding parameters used in the calculation for both phases are
summarized in Table~\ref{tb-params}.

For million-atom systems the size of the tight-binding
Hamiltonian\cite{jancu,zielinski-including} typically exceeds $10^7$,
reaching $10^8$ for the largest crystal phase quantum dot considered in this
work.
However, due to the nearest-neighbor approximation, the Hamiltonian matrix
is sparse and the number of non-zero matrix-elements scales as $O(N)$.
Hamiltonian elements are calculated on-demand without occupying computer's
memory.
Several lowest electron and hole states are found by means of Lanczos
algorithm with matrix-vector multiplication parallelized using the {\em MPI}
library.
We reiterate that thanks to the application of Lanczos algorithm, and the
sparse form of Hamiltonian, the process of partial Hamiltonian
diagonalization scales linearly with the domain size.
More details will be shown in the following section.

\section{Benchmarks}
Once single-particle states are calculated, we move to the efficient
calculation of Coulomb matrix elements.
We first illustrate our method on the example of
``standard''\cite{Zunger-EPM1,zielinski-including} lens-shaped InAs/GaAs
quantum dot.
The quantum dot has a diameter of $25$~nm and a height of $3.5$~nm and is
located on a $0.6$~nm thick wetting layer.
The InAs quantum dot and the wetting layer are embedded into GaAs matrix,
with the total number of atoms in the tight-binding
computational domain reaching $0.6\times10^6$ atoms.
This ``standard'' quantum dot will be used to benchmark our approach, while
later in the paper we will present results for InP crystal phase quantum
dots.

In the tight-binding method the Hamiltonian matrix elements are treated as
empirical parameters and the basis is not explicitly specified.
Due to this freedom there are several practical choices possible for the basis
functions.
These typically include Slater orbitals,\cite{slater,lee-johnsson-prb2001}
and Hermann-Skillman orbitals.\cite{hs-orbs,malkova-bryant}
Slater-type orbitals are defined by simple rules giving approximate analytic
atomic wave functions, in form of $\phi_{\mu}=N_{\mu}
Y_{\mu}\left(\theta,\phi\right) r^{n-1}e^{-\alpha r}$, where $n$ is the
principal quantum number, $N_{\mu}$ is the normalization constant, $\mu$
denotes orbital symmetry, and $Y_{\mu}\left(\theta,\phi\right)$
corresponding spherical harmonic, finally $\alpha$ is a screening constant
obtained by a set semi-empirical principles.
On the other hand, Hermann-Skillman orbitals are be obtained by
self-consistent numerical calculations for free atoms and ions.
Importantly, we note here that there is ongoing research for new
tight-binding schemes\cite{voisin-fit,tan-fit} that would remove the basis
ambiguity and directly relate tight-binding parameters with a well defined
orbital set.

\begin{figure}
  \begin{center}
  \includegraphics[width=0.5\textwidth]{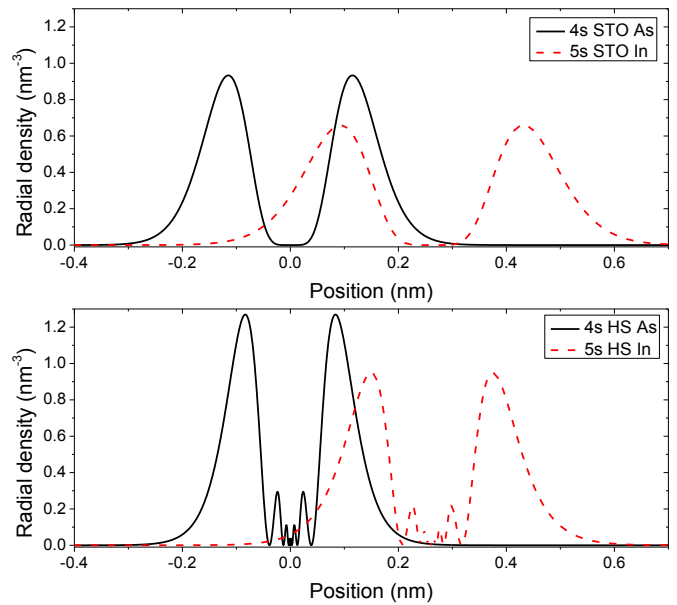}
  \end{center}
  \caption{Comparison of Slater-type (upper row) orbitals and Herman-Skillman (lower row) orbitals charge density distribution for indium and arsenic atomic $s$ valence orbitals. The distance between indium and arsenic corresponds to InAs bond length.}
  \label{hs-slater}
\end{figure}

Slater orbitals have analytical formulation, however Hermann-Skillman
orbitals should correspond more closely to the actual atomic states.
None of these basis sets is orthogonal (due to non-zero overlaps between
orbitals on neighboring atoms), as assumed by the Slater-Koster
tight-binding approach, however Hermann-Skillman orbitals are also better
localized in space (Fig.~\ref{hs-slater}) and have smaller overlap between
neighboring sites, whereas Slater orbitals have tails extending over many
lattice sites.

Our implementation accepts any form of orbitals defined as a product of
radial and angular parts,
whereas the radial part can be given by an analytical or numerical form.
Therefore our method can, in principle, utilize orbitals
such as those generated by TB-DFT approach\cite{tb-dft}, or new schemes of
empirical tight-binding parameterization.\cite{voisin-fit,tan-fit}

\begin{figure}
  \begin{center}
  \includegraphics[width=0.5\textwidth]{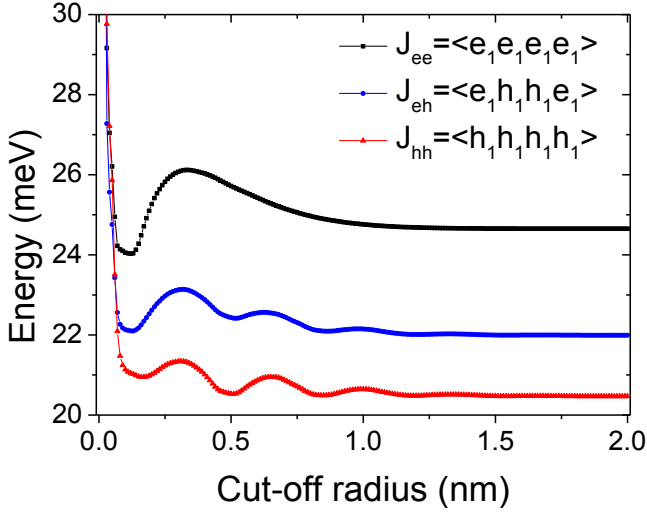}
  \end{center}
  \caption{Electron-electron $J^{ee}$, electron-hole $J^{eh}$ and hole-hole $J^{hh}$ Coulomb integrals for electron and hole occupying their ground states calculated for the lens-shaped (see the text) InAs/GaAs quantum dot as a function of the wave-function cut-off radius.}
  \label{cut-off}
\end{figure}

Figure~\ref{cut-off} shows electron-electron $J_{ee}$, electron-hole
$J_{eh}$ and hole-hole $J_{hh}$ Coulomb integrals for electron and hole
occupying their ground $s$ states\cite{zielinski-vbo},
calculated as a function of wave-function reconstruction cut-off radius
using Herman-Skillman basis, grid step $h=0.8$~\AA, and Thomas-Fermi
screening model for a lens-shaped InAs/GaAs quantum dot.
Here we use notation, where e.g.
$J_{eh} \equiv V_{e_1h_1h_1e_1} \equiv \langle\,e_1h_1h_1e_1\,\rangle$.
These three integrals are of particular importance as they allow to estimate
the single exciton and excitonic complexes binding energies at the level of
Hartree-Fock approximation.\cite{zielinski-vbo}
Figure~\ref{cut-off} shows that the cut-off radius of $\approx 1.5$~nm is
sufficient for the converged (within $0.1\%$ accuracy) calculation of these
Coulomb integrals.
We performed numerical tests for numerous quantum dot systems (including
nanocrystals, self-assembled and nanowire quantum dots) and generally found
that the cut-off radius of 1.5--2.0~nm for Herman-Skillman,
and 2.0--2.5~nm for Slater orbitals, is sufficient to achieve this level of
accuracy.
We also found that the cut-off radius is, to a large degree, independent
from the size and shape of the investigated nanosystem.

We note as well that there are $\approx 500$ atoms within $1.5$~nm cut-off
radius, potentially contributing to a single grid point.
As discussed earlier this is a much smaller number then total $\approx 10^6$
number of atoms in the computational domain.
Therefore without the $R_\text{cut}$ optimization (Eq.~\ref{tb-cut-off}) the
reconstruction stage would take $10^3$ times longer, rendering the entire
calculation impractical.

\begin{figure}
  \begin{center}
  \includegraphics[width=0.5\textwidth]{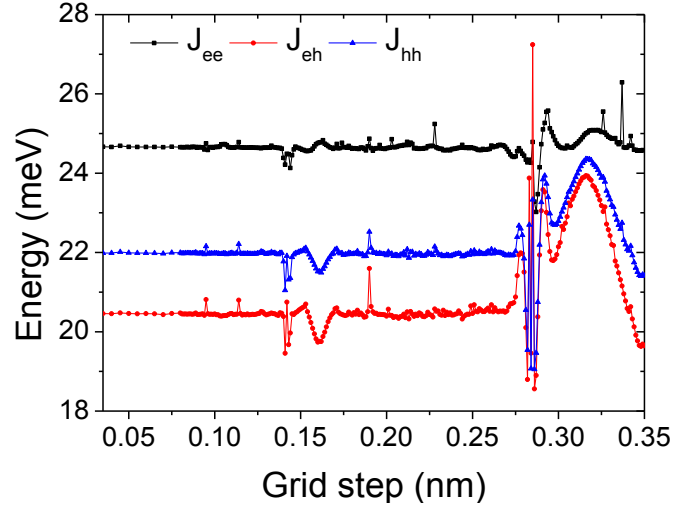}
  \end{center}
  \caption{Electron-electron $J^{ee}$, electron-hole $J^{eh}$ and hole-hole $J^{hh}$ Coulomb integrals for electron and hole occupying their ground states calculated for the the lens-shaped (see the text) InAs/GaAs quantum dot
  as a function of the wave-function reconstruction grid step.}
  \label{grid-h}
\end{figure}

Figure~\ref{grid-h} shows the evolution of the same Coulomb integrals for
the same quantum dot system as discussed earlier (using Herman-Skillman
orbitals as well) however as a function of the grid spatial resolution and
assuming a fixed cut-off radius equal to $1.5$~nm.
For large grid steps there are noticeable oscillations related to the
overlap of the regular computational grid and the underlaying zinc-blende
crystal lattice.
These oscillations are particularly pronounced at the grid spacings
0.2--0.3~nm comparable to typical InAs and GaAs bond-lengths, however for a
grid step $h$ lower than $0.09$~nm integral values stabilize.
Importantly, already for $h\approx 0.1$~nm the relative uncertainty due to
grid spacing errors is $<1\%$.
We checked that these conclusions can be generalized to all other integrals
entering the excitonic calculation.
Unless specified otherwise, the results presented below refer to the
$h=0.08\,\text{nm}=0.8\,\text{\AA}$ grid step.
Further increase of grid resolution (decrease of $h$) seem unnecessary as
excitonic properties of quantum dots are determined by contributions from
valence orbitals rather the subatomic details of atomic cores.

For $h\approx0.1$~nm there is about $100$ grid points per atom, each point
storing spin-up and spin-down part of the wavefunction.
For comparison, in the LCAO form and $sp^3d^5s^\star$ parameterization there
are $20$ spin-orbitals per atom and therefore $20$ (complex) expansion
coefficients per single atom.
Therefore we note the memory needed to store the wave-function in the
real-space representation is increased by factor $\approx 10$ (20 for
$sp^3s^\star$) when compared to the conventional TB-LCAO form.
For a million-atom system this corresponds to $\approx 3\,\text{GB}$ per
single TB wavefunction and can be efficiently handled by modern day
computers.
To summarize, the cut-off radius of about $2$~nm and grid-step of about
$0.1$~nm were found satisfactory for the reasonable convergence of selected
Coulomb matrix elements.

Table~\ref{tablecompare} shows various Coulomb direct and exchange integrals
calculated for InAs/GaAs lens-shaped quantum dot ($D=25$~nm, $h=3.5$~nm)
using different approaches.
\begin{table*}[tp]
\caption{Selected Coulomb and exchange integrals calculated for lens-shaped
(see the text) InAs/GaAs quantum dot for different basis and two different
models of dielectric screening: bulk-like $\epsilon_{InAs}$ screening and
Thomas-Fermi (T-F) model of Resta (see the text). Results obtained using
simplified atomistic model (TB-LCAO;~Eq.~\ref{coulomb-special-simple}) are
shown for comparison.}
\label{tablecompare}\centering
\begin{center}
   \begin{tabular}{ c | c c | c  c | c | c}
     & \multicolumn{2}{|c}{STO} & \multicolumn{2}{|c}{H-S orbitals} & \multicolumn{1}{|c}{TB-LCAO} & \multicolumn{1}{|c}{Opt. STO} \\
     & \small{$\epsilon_{InAs}$} & \small{T-F} & \small{$\epsilon_{InAs}$} & \small{T-F} & \small{$\epsilon_{InAs}$}  & \small{T-F} \\ \hline
    $J_{ee}=\langle\,e_1e_1e_1e_1\,\rangle$ [meV]     & 26.00  & 26.05  & 24.61  & 24.65  & 24.75  & 25.82\\
    $J_{eh}=\langle\,e_1h_1h_1e_1\,\rangle$ [meV]     & 20.03  & 20.06  & 21.96  & 21.99  & 22.54  & 22.66\\
    $J_{hh}=\langle\,h_1h_1h_1h_1\,\rangle$ [meV]     & 17.54  & 17.58  & 20.41  & 20.47  & 21.25  & 20.81\\ \hline
    $|\langle\,e_{\uparrow}h_{\Uparrow}  e_{\uparrow}  h_{\Uparrow}\,\rangle|$ [$\mu$eV] & 708.5  & 722.8  & 233.5  & 254.9  & 238.6  & 154.2\\
    $|\langle\,e_{\uparrow}h_{\Downarrow}e_{\downarrow}h_{\Uparrow}\,\rangle|$ [$\mu$eV] & 553.8  & 555.9  & 50.1  & 49.7  & 22.5  & 20.8 \\
    $|\langle\,e_{\uparrow}h_{\Uparrow}  e_{\downarrow}h_{\Downarrow}\,\rangle|$ [$\mu$eV] & 13.4  & 13.5  & 0.6  & 0.6  & 0.2  & 0.6 \\
   \end{tabular}
\end{center}
\end{table*}
For this particular nanosystem the electron-electron repulsion has larger
magnitude than electron-hole attraction or hole-hole repulsion.
In all considered cases there is only a minor difference due to dielectric
screening model used in a calculation.
This confirms our general conclusion that Coulomb interactions in typical
self-assembled quantum dots are nearly bulk screened as discussed earlier in
the text.
For the case of $J_{ee}$ all approaches give similar value of 25--26~meV.
There is, however, a substantial discrepancy
between different approaches for direct integrals involving hole states,
i.e.  $J_{eh}$ and $J_{hh}$.
This is particularly noticeable for $J_{hh}\approx 17.5$~meV obtained using
Slater-type orbitals, whereas both Herman-Skillman orbitals and the simple
model of Eq.~\ref{coulomb-special-simple} predict
$J_{hh}$ to vary between $20.5$ and $21.25$~meV.

The above differences can be understood in terms of atomic orbitals
contributions to single particle states.
Whereas, the ground electron and hole states have quite similar
envelopes\cite{zielinski-including}, they have much different ($s$- and
$p$-type correspondingly) dominant orbital contributions.
In particular, $p$-type (and $d$-type) Slater-type orbitals are typically
substantially more delocalized in space than corresponding Herman-Skillman
orbitals. Therefore the difference between both basis sets is due to spatial
extent of basis orbitals, what will be verified later in the text.

Apart from Coulomb direct integrals, Table~\ref{tablecompare} shows three
selected exchange integrals.
Those integrals play important role in the control of quantum dot excitonic
fine structure.\cite{bayer-eh}
The dark-bright exciton exchange splitting is determined predominantly by a
(real) exchange matrix element, which
also conserves spin:
$\langle\,e_{\uparrow}h_{\Uparrow}e_{\uparrow}h_{\Uparrow}\,\rangle$,
whereas
$\langle\,e_{\uparrow}h_{\Downarrow}e_{\downarrow}h_{\Uparrow}\,\rangle$ is
responsible for mixing of two bright-excitonic
states ($e_{\uparrow}h_{\Downarrow}$ and $e_{\downarrow}h_{\Uparrow}$) and
therefore leads to the bright-exciton splitting.
Finally
$\langle\,e_{\uparrow}h_{\Uparrow}e_{\downarrow}h_{\Downarrow}\,\rangle$
mixes two dark states ($e_{\uparrow}h_{\Uparrow}$ and
$e_{\downarrow}h_{\Downarrow}$)
and leads to the dark-exciton splitting.  In all above cases we used
notation where $e_{\uparrow}$ and $e_{\downarrow}$ is a Kramers degenerate
pair of states corresponding to the electron ground state energy.
Analogous notation has been used for hole states as well.

For the case of exchange integrals the effect of the basis has very
noticeable consequences.
Again, whereas results obtained using Herman-Skillman orbitals and
Eq.~\ref{coulomb-special-simple} agree reasonably, especially for the
electron-hole exchange ($\approx 0.25$~meV), the value calculated by using
Slater-type orbitals is about three time larger and close to $0.8$~meV.
Even larger difference is reported for integrals related to the
bright-exciton splitting (anisotropic electron-hole exchange) and the
dark-exciton splitting.
In this case the Slater-type orbitals basis overestimates the other two
approaches by more than an order of magnitude.
It should be noted that the bright-exciton splitting (fine structure
splitting) for cylindrical (or close to cylindrical) quantum dots observed
in the experiment is typically on the order of $10$ to $100$~$\mu$eV,
therefore in clear disagreement with the approach using Slater-type
orbitals.

\begin{figure}
  \begin{center}
  \includegraphics[width=0.5\textwidth]{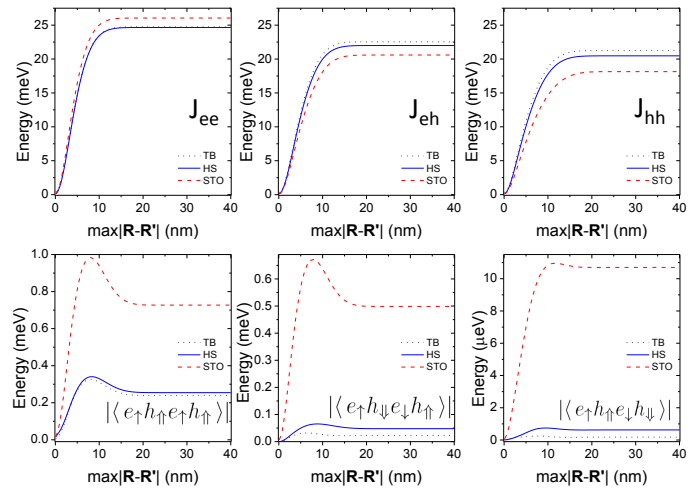}
  \end{center}
  \caption{Several Coulomb and exchange integrals calculated for lens-shaped (see the text) InAs/GaAs quantum dot as a function of a cut-off radius of Coulomb interaction.}
  \label{interaction-cutoff}
\end{figure}
To study this effect further, Figure~\ref{interaction-cutoff} shows the same
integrals as presented in Table~\ref{tablecompare}, however calculated as a
function of the Coulomb interaction radius which can be artificially limited
to a certain radius $r_\text{max}\equiv
\max\left|\mathbf{R}-\mathbf{R'}\right|$.
In this case for a simple model, Eq.~\ref{coulomb-special-simple} is simply
replaced with the following formula:
\begin{equation}
V_{nm} = \left\{
  \begin{array}{lcl}
    U_n &:& n=m \,,\\
    \frac{1}{\epsilon|\vec{R}_n-\vec{R}_m|} &:& n \ne m \quad\wedge\quad |\vec{R}_n-\vec{R}_m|\le r_\text{max} \,,\\
    0 &:& n \ne m \quad\wedge\quad |\vec{R}_n-\vec{R}_m|>r_\text{max} \,.
  \end{array}
\right.
\end{equation}

Whereas for grid calculations, in order to avoid granularity artifacts we
use the sigmoid-type function:
\begin{equation}
G(\vec{r}) = \frac{e^2}{4\pi\epsilon} \frac{1}{|\vec{r}|}
\left( 1+e^{ \frac{1}{\lambda} (|\vec{r}|-r_\text{max}) } \right)^{-1} \,,
\end{equation}
where $\lambda>0$ plays a role of a smoothing parameter.
For clarity we emphasize that the Coulomb interaction cut-off radius is a
different quantity from the wave-function reconstruction cut-off radius
studied in Fig.~\ref{cut-off}.

For all considered integrals their values stabilize for the Coulomb
interaction cut-off radius $\approx 20$~nm comparable to the quantum dot
diameter. This further proves that the calculation is free from any effects
of the image charges.
Fig.~\ref{interaction-cutoff} (upper row) demonstrates rather small
contributions to the direct matrix elements for this nanosystem that
originate from on-site and nearest-neighbor contributions, and that these
integrals are dominated by the long-range contributions.
Interestingly, for the case of exchange integrals there are noticeable
maxima in their modulus at about 8--10~nm, corresponding to approximately
one-third of quantum dot diameter (or three times the quantum dot height).
With further increase of the Coulomb interaction cut-off radius modules of
exchange integrals are reduced and their values stabilize at $20$~nm.
As this effect is visible for the simple model
(Eq.~\ref{coulomb-special-simple}) as wells as for calculations involving
basis, this suggest that there are two long-range,
monopole-monopole contributions to exchange integrals of opposite character.
On the other hand, for the case of exchange integrals calculated using
Slater-type orbitals the short-range ($<3$~nm) contribution is apparently
dominant and is likely responsible
for the overestimation of integral values as compared to other approaches.

As discussed earlier, in case of semiconductor quantum dots, one could
assume that for sites which are far enough apart from each other the exact
structure of the localized orbitals is
not important\cite{schulz-schum-czycholl} and that the long-range
contributions are dominated by the monopole-monopole interaction of two
charge densities localized at different sites.
This assumption is clearly not fulfilled for Slater-type orbitals which
extend over many lattice constants from the site center.
This is especially important for the bright-exciton splitting that has been
shown to be strongly related to the (local) electron-hole non-orthogonality
on the scale of a unit cell.\cite{Franceschetti,Goupalov}
Slater-type orbitals localized on neighboring sites, i.e. within the same
unit cell, are clearly (Fig.~\ref{hs-slater}) far from orthogonal, resulting
in strongly overestimated value of
anisotropic exchange integral.

\begin{figure}
  \begin{center}
  \includegraphics[width=0.5\textwidth]{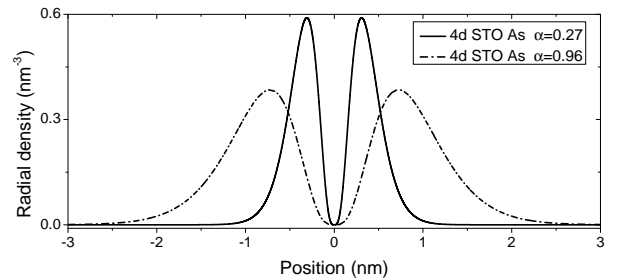}
  \end{center}
  \caption{Comparison of radial densities of Slater-type orbitals for arsenic $4d$ orbital, with original ($\alpha=0.27$) and modified ($\alpha=0.96$; see the text) screening constants.}
  \label{sto-compress}
\end{figure}

In order to study these effects further, we note that bulk on-site atomic
energies are shifted with respect to their free atom
counterparts.~\cite{jancu}
By the same token, the spatial extent of basis orbitals in bulk should be
reduced as compared with free atoms.
This claim has been recently supported by work by Benchamekh et
al.~\cite{voisin-fit} where microscopic (Bloch) functions for tight-binding
model where obtained by a process in which
screening constants $\alpha$ of Slater-type orbitals were optimized by the
fitting.
Ref.~\cite{voisin-fit} shows that the fitting procedure has relatively small
effect on well localized $4s$ and $4p$ orbitals ($sp^3$ in tight-binding),
e.g.  altering $4s$ screening constant from $1.7$ to $1.94$.
On the other hand, the screening constant of Slater-type $4d$ arsenic
orbital is increased significantly from $0.27$ in free atom case to $0.96$
for arsenic site in GaAs bulk crystal.
The screening constant of $5s$ (i.e.  $s^\star$) orbital is affected even
more, being increased from $0.4$ to over $1.74$ for the bulk case.
This effectively corresponds to effective ``compression'' of atomic orbitals
in bulk, and is especially pronounced for higher orbitals of the largest
spatial extent.
These results indicate strong (exponential) dumping of atomic orbital tails
at long distance as compared to free atoms counterparts, what is apparent
radial density plot as seen on Figure~\ref{sto-compress}.
This procedure reduces significantly basis functions tails and the
contribution from a given site is on average effectively limited to the
radius of about $1$~nm.

Whereas modified screening constants are currently not available for indium
arsenide, we repeated our calculations using an approach in which we
increased screening constants on $s$, $p$, $d$, and $s^\star$ orbitals all
atomic species in the nanosystem by the following values: $0.36$, $0.25$,
$0.7$, $1.3$.
These values are close to modifications reported by Ref.~\cite{voisin-fit}.
Results of our calculations are shown on the right-hand column (Opt.  STO)
in Table~\ref{tablecompare}.

In the optimized case Coulomb and exchange integral values are similar to
those given by a more confined Herman-Skillman basis or the asymptotic model
of Eq.~\ref{coulomb-special-simple} (TB-LCAO).
This is particularly pronounced for exchange integrals, e.g.  for
$\left|\langle\,e_{\uparrow}h_{\Downarrow}e_{\downarrow}h_{\Uparrow}\,\rangle\right|$
(related to the bright-exciton splitting) where
the ``compressed'' Slater-type orbitals give values order of magnitude
smaller than unmodified orbitals and very close to that of
Eq.~\ref{coulomb-special-simple}.
Otherwise, we note that the choice of unmodified Slater-type basis results
in severe overestimation of above integrals due to the overlap of orbitals
between far ($>1.0$~nm) neighbors.

\section{Crystal phase quantum dots}
\begin{figure}
  \begin{center}
  \includegraphics[width=0.5\textwidth]{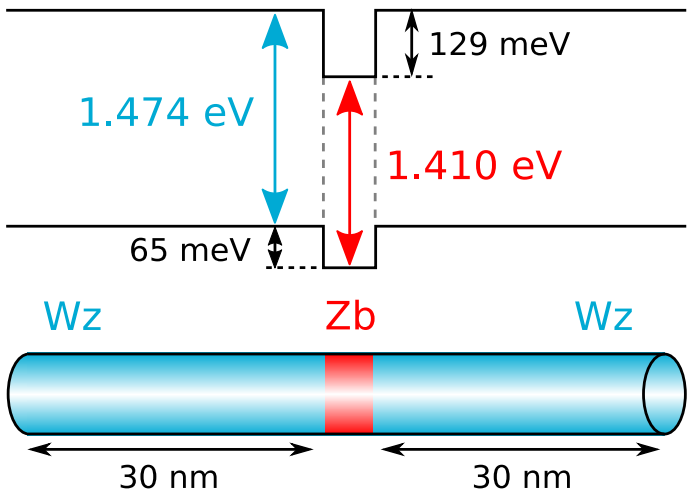}
  \end{center}
  \caption{Band alignment of InP zinc-blend and wurtzite used in our calculations, and schematics of the system.}
  \label{band-aligment}
\end{figure}

Crystal phase quantum dots\cite{nika-cpqd,our-cpqd,zhang-cpqd} gained
recently a lot of attraction due to nearly perfect interface between crystal
phases constituting the system, well defined (with monolayer accuracy)
height, and the lack of alloying effects pestering spectral
reproducibility\cite{alloyed-qd,Mlinar-inverse,Zielinski-natural} of typical
self-assembled quantum dots.  In this section we apply our method to a
single InP crystal phase quantum dot.
We model this nanostructure by a zinc-blende InP segment of 1 nm height (3
monolayers, single ABC stacking sequence) along the [111] direction,
embedded between two 30 nm (100 MLs) long wurtzite InP segments grown along
$[0001]$ direction.  Fig.~\ref{band-aligment} shows corresponding energy
band alignment of the studied system, plotted along the growth axis, and
schematics of the nanosystem.
The total height of the system exceeds $60$~nm, whereas we vary nanowire
diameter from $12$ to $70$ nm, therefore a number of atoms in the system
varies from close 0.3 to over 10 million atoms.

\begin{figure}
  \begin{center}
  \includegraphics[width=0.5\textwidth]{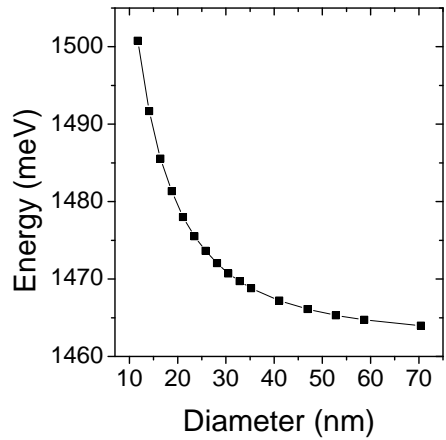}
  \end{center}
  \caption{Single particle effective gap calculated for crystal phase quantum dots of different diameters (see the text).}
  \label{tb-gap}
\end{figure}
Fig.~\ref{tb-gap} shows the effective single particle gap (defined as the
energy difference between ground electron and hole state) calculated for the
crystal phase quantum dot as a function of a diameter.
Importantly, the effective gap varies substantially from over $1.5$~eV for
the lowest considered diameter ($11.7$~nm) to $\approx 1.46$~eV for the
largest considered diameter of $70.4$~nm.
This results shows that it is in principle possible to tailor the effective
gap of these nanostructure by control of a nanowire diameter.
On the other hand, Fig.~\ref{tb-gap} reveals the significant role of lateral
confinement, therefore without precise control of diameter crystal phase
quantum dots are subject to a potential spectral inhomogeneity.

\begin{figure}
  \begin{center}
  \includegraphics[width=0.5\textwidth]{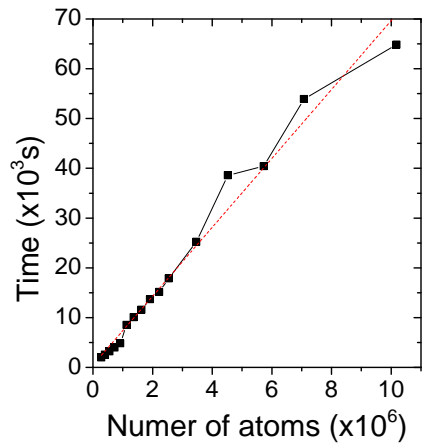}
  \end{center}
  \caption{Time of the tight-binding (singe particle) calculations for crystal phase quantum dots (see the text)
  of different number of atoms (quadratically dependent on the nanostructure diameter);
  $10.2$ millon atoms corresponds to $70.4$~nm nanometers of diameter.
  The red (dashed) line is a linear fit that is a guide to the eye.}
  \label{tb-time}
\end{figure}
Fig.~\ref{tb-time} shows the time of the single particle (tight-binding)
part of calculation as function of number of atoms (which is growing
quadratically with the nanowire diameter).
Each of the computations were performed on the same $48$-core computer
system. The time of the computation scales practically linearly as a function
of number of atoms.
The small steps on the plot are related to parallel computation and
load-balancing issues, i.e.  problem of uniform division of a discrete
atomic grid to a discrete number of processors.

As mentioned in the introduction, experimental spectra are obtained for the
interacting electron-hole pair, i.e.  exciton.
Therefore we follow our single-particle calculation with a many-body
configuration interaction (CI) calculation.
We performed our calculations using two approaches, first is the traditional
TB-LCAO method of Eq.~\ref{coulomb-special-simple} that involved $O(N^2)$
summations, and the second is our linear-scaling approach.
In other to execute this comparison we have utilized a limited CI basis
involving lowest three (with spin, six) electron states and lowest four
(with spin, eight) hole states.
For each diameter, we have calculated the total of $1296$ electron-electron
integrals, $2\times2304=4608$ electron-hole Coulomb and exchange integral,
and $4096$ hole-hole integrals.
{\footnotesize Note: Number of integrals to be calculated is effectively
reduced by a factor of four due to the symmetries of Coulomb matrix
elements.}
Computation of these integrals is a prerequisite not only for the single
exciton calculations (which involve only electron-hole interactions),
but also for other excitonic complexes such as a biexciton (that
additionally needs electron-electron and hole-hole integrals).
In particular, the above basis results in total $48$ configurations for the
single exciton and $420$ configurations for the biexciton.
Such number of configurations presents little computational effort for the
CI Hamiltonian diagonalization;
however, due to rapidly ($M^4$) increasing number of Coulomb matrix
elements, further increase of the CI basis would results in prohibitive
computational times
for the method of Eq.~\ref{coulomb-special-simple}, and would render the
comparison between methods impractical.

\begin{figure}
  \begin{center}
  \includegraphics[width=0.5\textwidth]{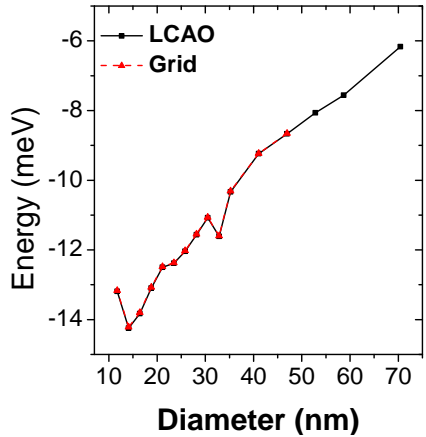}
  \end{center}
  \caption{Excitonic binding energy (see the text) calculated for crystal phase quantum dots of different diameters (see the text).}
  \label{X-binding}
\end{figure}

Fig.~\ref{X-binding} shows the excitonic binding energy for the discussed
crystal phase quantum dot as a function of diameter.
The exciton binding energy is defined as a energy difference between the
ground state of the interacting electron-hole pair and the single particle
gap studied earlier.
With the increasing of diameter, the magnitude (absolute value) of the
binding energy decreases from a high-lateral confinement regime ($\approx
14$~meV) for small diameters, to nearly bulk-like binding
energy of $\approx 6$~meV for the largest diameter.  This figure reveals a
transition from quantum dot-like to a quantum well-like confinement for
large diameter systems.
The unexpected steps (e.g.  $33$~nm) in otherwise monotonous function are an
artefact of a limited (not converged) configuration interaction basis set.
Most importantly, we note the in the range where both methods can be
compared with each other ($d<50$~nm) they produce practically identical
output. We note as well the results shown here were obtained for the grid
step of $1.0$~\AA, yet they differ only by a tiny fraction of a meV
(typically $0.05$~meV) from results obtained on the $0.8$~\AA~grid.
Additionally, we point that due to $C_{3v}$ symmetry crystal
phase quantum dots have exactly vanishing excitonic fine structure\cite{singh-bester-eh} as
confirmed by our studies, therefore these systems are to a large degree
free from basis dependence artifacts mentioned earlier.

\begin{figure}
  \begin{center}
  \includegraphics[width=0.5\textwidth]{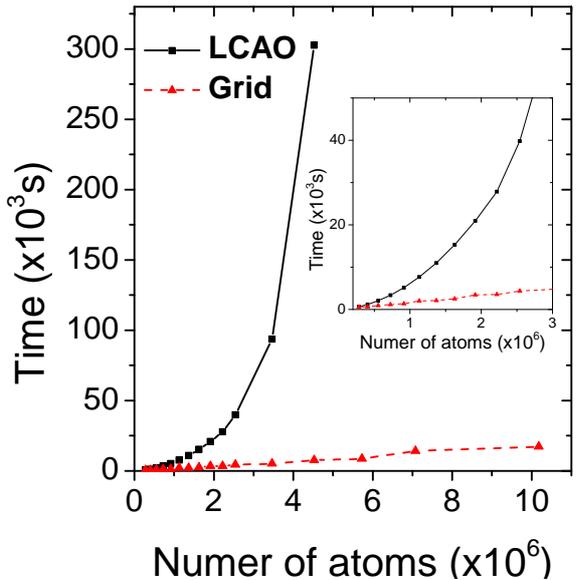}
  \end{center}
  \caption{Time of the Coulomb matrix elements calculation for two different approaches (see the text).}
  \label{CME-time}
\end{figure}
Whereas the excitonic results reported by both methods are nearly identical,
the cost of the computational differs dramatically as shown on
Fig.~\ref{CME-time}.
With the domain size growing from $0.28$ to $4.5$ million atoms, the time
spend in Eq.~\ref{coulomb-special} calculation is increased by more than
$\approx 460$ times, showing strong non-linear scaling.
This scaling is even more unfavorable than $O(N^2)$, most likely due to the
increased memory-bandwidth usage (and bandwidth bottleneck) for the large number of atoms.
On the other hand, the wave-function reconstruction approach shows nearly
linear scaling, i.e.  the time of computations is increased by $\approx 30$
factor, corresponding to similar growth of the number of atoms in the
domain. The timings presented on Fig.~\ref{CME-time} were obtained for grid
spacing equal to $1$~\AA. We note that for higher resolution grids the time
and memory demand is increased, e.g. by $\approx 40\%$ factor for the $0.9$~\AA
grid spacing.

\begin{figure}
  \begin{center}
  \includegraphics[width=0.5\textwidth]{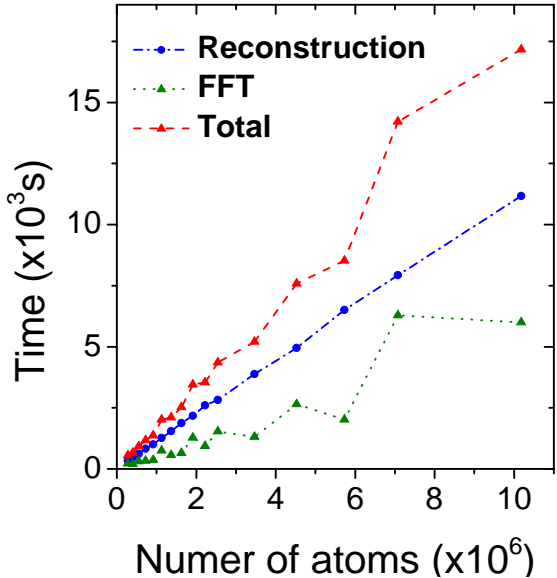}
  \end{center}
  \caption{Times of the grid reconstruction and the FFT stages (see the text) versus the total time of the computation.}
  \label{CME-fft-time}
\end{figure}
In principle, a further speedup could be achieved: both results were
obtained on the same $240$-core computer cluster system, however the FFT
part of the calculation where run on 48~cores only, due to the limited
scaling of the FFT algorithm (as implemented in the FFTW library, which we
have used in this work).  Artifacts of the FFT parallelization are well
visible on Fig.~\ref{CME-fft-time} where the total time of the calculation
in our approach was divided into the contribution from the reconstruction
part and the FFT step of the computations.
The reconstruction itself scales practically linear as function of number of
atoms and in fact the overall time is dominated by this stage.
Noticeable steps in time dependence of the FFT part are related to a)
additional padding of the reconstruction domain, to ensure good performance
of the FFT, and b) the load balancing issues on parallel computer system.
In fact, the FFT time spent in $10$-million case is smaller than in
$7$-million case, as the grid in the larger case can be more efficiently
divided by the number of processors.

We have checked that quasi-linear scaling properties of our approach are not
limited to this particular system, but instead, it is a general feature,
with similar scaling properties for various nanosystems such as:
nanocrystals, nanowires, self-assembled and nanowire quantum dots.

\section{Summary}
We have presented a method for an efficient, \mbox{order-$N$} calculation of
excitonic spectra of semiconductor nanosystems.
Our method is directly applicable to ten-million-atom nanostructures, such
as crystal phase quantum dots.
The first steps of strain (only for lattice mismatched systems) and the
empirical tight-binding calculation are followed by the efficient
calculation of Coulomb matrix elements and the configuration interaction
approach.
Our approach redefines the tight-binding LCAO single particles
wave-functions on a three-dimensional, regular grid.
The process of wave-function reconstruction can be performed efficiently by
relying of a finite extent of basis orbitals, with the cut-off radius
smaller than 2.5~nm.
The grid step of about 0.8--1.0~\AA was found sufficient for the convergence
of the key Coulomb and exchange integrals affecting excitonic spectra.
The grid reproduced form of the wave-function allows for the application of
Fourier space methods for the calculation of Coulomb matrix elements that
benefit tremendously from the Fast Fourier Transform algorithm.
Our method goes beyond traditional two-center and monopole-monopole
approximations, further our approach can account for different models of
dielectric screening function and utilize different basis set such as those
generated by TB-DFT approach or new schemes of tight-binding
parameterization.
We studied the role of a basis by comparing results obtained with
Slater-type and Herman-Skillman orbitals and the simplified $O(N^2)$
approach.
We analyzed the effects of basis locality and orthogonality and found that
long-range tails of basis orbitals affect significantly values of important
electron-hole exchange integrals.
The effect of local basis orthogonality is particularly important for
calculations involving bright- and dark- excitons splitting.
The straightforward application of highly non-orthogonal Slater-type
orbitals would lead to severe overestimation of bright-exciton splitting as
compared with other approaches.
On the other hand Herman-Skillman orbitals, or recently optimized
Slater-type orbitals, should correspond more closely to the actual atomic
states, however more accurate calculations of the excitonic fine structure
should involve basis sets directly designed for tight-binding calculations.

Finally, we illustrate our approach with the calculation for the crystal
phase quantum dots with diameter exceeding 70~nm and number of atoms
exceeding 10~million.
We demonstated pronounced diameter dependence of both the effective gap and the
excitonic binding energy, with transition to nearly bulk-like binding energy
for the largest diameter system.

Linear scaling with number of atoms opens an route for semi-empirical
atomistic calculations of large semiconductor systems with number of atoms
reaching $10^8$ in near future.
Due to its efficiency, the method should allow for new possible application
such as self-consistent calculations for multi-million-atom nanosystems or
accurate nanostructure excitonic calculations involving large many-body
basis sets.  The range of potential applications varies from
quantum dots of different species, through nanowires, up to modeling of
electronic properties of single dopants embedded in multi-million atoms
transistors.

\section{Acknowledgment}
Support from the Foundation for Polish Science, Homing Plus Programme,
co-financed by the European Union within the European Regional Development
Fund is kindly acknowledged.
M.Z.  acknowledges support from the Polish Ministry of Science and Higher
Education as a research project No IP 2012064572 (Iuventus Plus).

\bibliography{LinearScaling}

\end{document}